\documentclass[11pt]{article}

\def\lb{\label}
\def\bb{\bibitem}
\def\be{\begin{equation}}
\def\ee{\end{equation}}
\def\ba{\begin{eqnarray}}
\def\ea{\end{eqnarray}}

\def\ol{\overline}
\def\rd{{\rm d}}

\def\nn{\nonumber}
\def\D{(D)}
\def\5{(5)}
\def\4{_{(4)}}
\def\2{\sqrt{2}}
\def\3{\sqrt{3}}

\def\R{{\cal R}}

\def\La{\ol\Lambda}
\def\z{\zeta}

\begin{document}
\begin{titlepage}
\title{\begin{flushright}\begin{small}    LAPTH-068/13
\end{small} \end{flushright} \vspace{2cm}
Bertotti-Robinson solutions in five-dimensional quadratic gravity}
\author
{G\'erard Cl\'ement\thanks{Email:gerard.clement@lapth.cnrs.fr}\\
\small{LAPTh, Universit\'e de Savoie, CNRS, 9 chemin de Bellevue,} \\
\small{BP 110, F-74941 Annecy-le-Vieux cedex, France}}

\date{December 14, 2013}
\maketitle

\begin{abstract}
We construct new solutions of five-dimensional quadratic gravity as
direct products of a constant curvature two-surface with a solution
of three-dimensional new massive gravity with constant scalar
curvature. These solutions could represent near-horizon limits of
five-dimensional asymptotically flat black strings or black rings. A
number of these non-asymptotically flat solutions are themselves
black strings or rings. A by-product of our analysis is the
construction of new solutions of four-dimensional quadratic gravity
obtained by toroidal reduction of the five-dimensional solutions
with flat transverse space. These again include black strings or
rings, and an $AdS_2\times T^2$ solution of $f(R)$ gravity for a
specific relation between the model parameters.
\end{abstract}

\end{titlepage}\setcounter{page}{2}

\section{Introduction}

It is well known that Einstein's theory of gravity does not lead
upon quantization to a perturbatively renormalizable theory. The
addition to the linear Einstein-Hilbert action of curvature squared
terms leads to a renormalizable theory \cite{stelle}. This theory is
generically non-unitary due to the occurreence in the linearized
theory of massive spin-2 ghost modes, which are absent only if the
quadratic action is of the Einstein-Gauss-Bonnet form \cite{zwi}.
However, it has recently been shown that in four \cite{crit} and in
higher dimensions \cite{crit2} the massive spin-2 mode may be
rendered massless by a (more general) suitable choice of the model
parameters.

Few exact solutions of the generic theory of $D$-dimensional
quadratic gravity are known, with the exception of classes of
algebraically special solutions, such as AdS waves \cite{ggst}, and
type III and type N solutions \cite{malpra} (for a review and
references, see \cite{malek}). This paper is devoted to the search
for exact solutions of five- (and four-) dimensional quadratic
gravity describing spacetimes which are the direct product of two
manifolds, one of which at least has constant curvature.

The first example of such a solution is due to Nariai \cite{nar},
who showed that the four-dimensional Einstein equations with a
positive cosmological constant admit a solution with the geometry
$dS_2\times S^2$. This has been shown to be the near-extreme,
near-horizon limit of the de Sitter-Schwarzschild solution
\cite{gp}. The anti-Nariai solution of the Einstein equations with a
negative cosmological constant, with the geometry $AdS_2\times H^2$,
can be generated from the Nariai solution by an appropriate duality
transformation \cite{dad}. Better known is the Bertotti-Robinson
solution \cite{br} of four-dimensional Einstein-Maxwell theory,
representing a spacetime with the geometry $AdS_2\times S^2$
supported by a monopole electric or magnetic flux. This is the
near-horizon limit of the extreme Reissner-Nordstr\"om black hole,
and can be generalized to a solution of the Einstein-Maxwell
equations with cosmological constant \cite{br} (for a review, see
\cite{lemos}).

The anti-Nariai and Bertotti-Robinson solutions have been
generalized to solutions of the $D$-dimensional cosmological
Einstein equations with the geometry $AdS_2\times \Sigma_{D-2}$,
with $\Sigma_{D-2}$ a constant curvature Riemannian manifold or a
product of such manifolds \cite{cvz,lemos2,KL8,KL10}. These are the
near-horizon limits of $D$-dimensional black holes. Near-horizon
geometries of extremal black hole solutions to higher-dimensional
supergravities were classified in \cite{KL13}. It has been shown
that such product spacetimes also solve the $D$-dimensional
Einstein-Gauss-Bonnet equations \cite{dad2} and, more generally, the
$D$-dimensional Einstein-Maxwell-dilaton equations with all possible
higher order corrections \cite{gurses}. In this paper, we shall
consider a different generalization to $D=5$ of the
Bertotti-Robinson solution, namely product spacetimes with the
geometry $M_3\times\Sigma_2$, where $M_3$ is a (not necessarily
constant curvature) Lorentzian manifold, and $\Sigma_2= S^2$, $T^2$
or $H^2$ is constant curvature. These could represent near-horizon
limits of five-dimensional black strings or black rings.

To construct such solutions to five-dimensional quadratic gravity,
we shall follow the strategy initiated in \cite{reds2}. The direct
product ansatz splits the five-dimensional field equations into a
longitudinal set of three-dimensional field equations (for the
$M_3$), and a transverse set of two-dimensional field equations (for
the $\Sigma_2$), which in the case of a constant curvature
$\Sigma_2$ amount to a constraint on the curvature invariants of the
$M_3$. This reduction is carried out in Section 2. It turns out that
in the present case the three-dimensional field equations are those
of new massive gravity (NMG) \cite{bht}, while the constraint
equation can be solved for a constant scalar curvature. In Section 3
we review the various known solutions of NMG with constant scalar
curvature, and the resulting uplifted solutions to five-dimensional
quadratic gravity. Depending on the values of the coupling
constants, these include in particular black strings and black
rings. In the special case of a flat transverse space $\Sigma_2$,
these five-dimensional solutions can be toroidally reduced to a
class of non-asymptotically flat cylindrical solutions of
four-dimensional quadratic gravity with the geometry $M_3\times
S^1$, again including some black strings or black rings. Our results
are summarized in the last section.

\setcounter{equation}{0}
\section{Reduction of five-dimensional quadratic gravity}

The generic action for $D$-dimensional quadratic gravity is
\cite{dt}
 \be I_{\D} = \int \rd^D\, x
\sqrt{|g_{(D)}|} \Big \{ \frac{\R-2\Lambda}{\kappa} + \alpha \R^2 +
\beta \R^{\mu\nu}R_{\mu\nu} + \gamma
(\R^{\mu\nu\rho\sigma}\R_{\mu\nu\rho\sigma} -4\R^{\mu\nu}\R_{\mu\nu}
+ \R^2 ) \Big \}. \label{action}
 \ee
where $\kappa$ is the $D$-dimensional Einstein constant, $\Lambda$
the cosmological constant and $\alpha$, $\beta$, $\gamma$ the
quadratic coupling constants.

The equations of motion that follow from (\ref{action}) are
 \ba &&
{1\over \kappa}(\R_{\mu \nu} -{1\over 2}g_{(D)\mu \nu} \R + \Lambda
g_{(D)\mu \nu}) + 2\alpha \R\,(\R_{\mu\nu} - {1\over
4}g_{(D)\mu\nu}\,\R ) \nn\\ && +2\beta (\R_{\mu \sigma \nu \rho} -
{1\over4}g_{\5\mu\nu}\,\R_{\sigma\rho})\R^{\sigma\rho} + 2\gamma
\Big \{ \R_{\mu\sigma\rho\tau}{\R_\nu}^{\sigma\rho\tau} -
2\R_{\mu\sigma\nu\rho}\R^{\sigma \rho} \nn\\ &&  - 2\R_{\mu
\sigma}{\R_\nu}^\sigma + \R\R_{\mu \nu} - {1\over 4}g_{\5\mu
\nu}(\R_{\tau\lambda\rho\sigma}\R^{\tau\lambda\rho\sigma} -
4\R_{\sigma\rho}\R^{\sigma\rho} + \R^2 )
\Big \} \nonumber \\
&& +(2\alpha +\beta) (g_{(D)\mu\nu}\nabla^2- \nabla_\mu \nabla_\nu
)\R +\beta \nabla^2 ( \R_{\mu\nu} - {1\over 2}g_{(D)\mu\nu} \R) = 0,
\label{eom}
 \ea
where ${\R^\mu}_{\sigma\nu\rho}$ and $\nabla_\mu$ are the Riemann
tensor and covariant derivative for the $D$-dimensional metric
$g_{(D){\mu\nu}}$.

Taking $D=5$, let us carry out dimensional reduction of the field
equations (\ref{eom}) relative to a constant curvature two-surface
$\Sigma_2$, assuming the direct product ansatz
 \be\lb{ans2}
\rd s_{\5}^2 = g_{\alpha\beta}(x^{\gamma})\rd x^{\alpha}\rd
x^{\beta} + a^2(\rd\theta^2 + s_k^2\rd\varphi^2)\,,
 \ee
where $\alpha,\beta,\gamma=1,2,3$, and $k=1,0$ or $-1$, with
 \be
s_1 = \sin\theta\,, \quad s_0 = \theta\,, \quad s_{-1} = \sinh\theta
 \ee
($\theta\in [0,\pi]$ for $k=1$ and $\theta\in [0,\infty]$ for
$k=0,\,-1$). The non-vanishing components of the Riemann tensor are
 \ba
{\R^{\alpha}}_{\beta\gamma\delta} &=& \delta^\alpha_\gamma
R_{\beta\delta} - \delta^\alpha_\delta R_{\beta\gamma}  +
g_{\beta\delta}{R^\alpha}_\gamma - g_{\beta\gamma}{R^\alpha}_\delta -
\frac12(\delta^\alpha_\gamma g_{\beta\delta} - \delta^\alpha_\delta
g_{\beta\gamma})R\,,\nn\\
{\R^{\varphi}}_{\theta\varphi\theta} &=& k\,, \quad
{\R^\theta}_{\varphi\theta\varphi} = ks_k^2\,,
 \ea
where $R_{\alpha\beta}$ is the three-dimensional Ricci tensor. The
five-dimensional and three-dimensional Ricci scalars are related by
$\R = R + 2ka^{-2}$. The only non-vanishing components of the
Lanczos tensor (the $2\gamma$ term in (\ref{eom})) are those of the
Einstein tensor:
 \be
L_{\alpha\beta} = 2ka^{-2}\left(R_ {\alpha\beta} - \frac12Rg_
{\alpha\beta}\right).
 \ee
Therefore, the five-dimensional Lovelock-Gauss-Bonnet equations
($\alpha=\beta=0$ in (\ref{eom})) reduce to the three-dimensional
cosmological Einstein equations with renormalized coupling
constants.

The ($\alpha\beta$) components of the full equations (\ref{eom}) reduce to
 \ba\lb{eom3}
&& \frac{\beta}2\,K_{\alpha\beta} +
\left(2\alpha+\frac{3\beta}4\right)(g_{\alpha\beta}D^2 -
D_{\alpha}D_{\beta})R \nn\\ && +
\left[\frac1{\kappa}+\frac{4k(\alpha+\gamma)}{a^2}+\left(2\alpha-\frac{\beta}4
\right)R\right]\left(R_{\alpha\beta}-\frac12g_{\alpha\beta}R\right)
\\ &&+ \left[\frac{\Lambda-ka^{-2}}{\kappa}-\frac{2k^2\alpha}{a^4}+
\left(
\frac{\alpha}2+\frac{11\beta}{16}\right)R^2\right]g_{\alpha\beta} =
0\,, \nn
 \ea
where
 \ba
K_{\alpha\beta} &=& 2D^2R_{\alpha\beta} -
\frac12(D_{\alpha}D_{\beta} + g_{\alpha\beta}D^2)R - 8
{R_{\alpha}}^{\gamma}R_{\beta\gamma} \nn\\ && +
\frac92RR_{\alpha\beta} + \left(3R^{\gamma\delta}R_{\gamma\delta}
-\frac{13}8R^2\right)g_{\alpha\beta}\,,
 \ea
and $D_{\alpha}$ is the three-dimensional covariant derivative. The
solutions of the three-dimensional field equations (\ref{eom3}) are
restricted by the constraint, resulting from the $\theta\theta$
equation (\ref{eom})
 \be\lb{cons}
\left(2\alpha+\frac{\beta}2\right)D^2R +
\left(\frac{\Lambda}{\kappa} + \frac{k^2(2\alpha+\beta)}{a^4}\right)
- \frac{1}{2\kappa}R - \frac{\alpha}2R^2 - \frac{\beta}2
R^{\gamma\delta}R_{\gamma\delta} = 0\,.
 \ee

Let us make the further assumption $R=$ constant. In this case, the
field equations (\ref{eom3}) reduce to the equations of new massive
gravity
 \be\lb{nmg}
R_{\alpha\beta}-\frac12g_{\alpha\beta}R + \lambda g_{\alpha\beta} -
\frac1{2m^2}K_{\alpha\beta} = 0\,,
 \ee
with the additional constraint
 \be\lb{cons1}
\frac{\beta}2R^{\gamma\delta}R_{\gamma\delta} +
\frac{\alpha}2R^2 + \frac{1}{2\kappa}R -
\left(\frac{k^2(2\alpha+\beta)}{a^4} + \frac{\Lambda}{\kappa}\right)
= 0\,.
 \ee
The parameters of the effective three-dimensional theory (\ref{nmg})
are related to the parameters of the original five-dimensional
theory, the compactification scale $a^2$ and the three-dimensional
Ricci scalar by
 \ba\lb{paras}
\lambda &=&
c^{-1}\left[\frac{\Lambda-ka^{-2}}{\kappa}-\frac{2k^2\alpha}{a^4}+
\left( \frac{\alpha}2+\frac{11\beta}{16}\right)R^2\right]\,, \nn\\
m^{-2} &=& -c^{-1}\beta\,, \quad c =
\frac1{\kappa}+\frac{4k(\alpha+\gamma)}{a^2}+\left(2\alpha-\frac{\beta}4
\right)R
 \ea
(the real constant $m^2$ is not necessarily positive). When
$\beta=0$ the reduced field equations (\ref{nmg}) reduce to the
three-dimensional cosmological Einstein equations, so it is the
five-dimensional quadratic Ricci coupling
$\beta\R^{\mu\nu}R_{\mu\nu}$ which is responsible for the occurrence
of the three-dimensional quadratic couplings in (\ref{nmg}).

Tracing the NMG field equations (\ref{nmg}) yields the equation
 \be
-\frac12R + 3\lambda -
\frac1{2m^2}\left(R^{\gamma\delta}R_{\gamma\delta} -
\frac38R^2\right) = 0\,,
 \ee
which can be rewritten as
 \be
\frac{\beta}2R^{\gamma\delta}R_{\gamma\delta} - \frac{3\beta}{16}R^2
- \frac{c}2R + 3c\lambda = 0\,.
 \ee
This can be combined with (\ref{cons1}) to eliminate the Ricci
square term. Taking (\ref{paras}) into account, we finally arrive at
the quadratic constraint on the Ricci scalar, equivalent to
(\ref{cons1}),
 \be\lb{cons2}
2\beta\ol{R}^2 - [1+2(\alpha+\gamma)x]\ol{R} + 4\ol{\Lambda} - 3x -
(4\alpha-\beta)x^2 = 0\,,
 \ee
where we have put $$\ol{R} = \kappa R\,, \quad \ol{\Lambda} =
\kappa\Lambda\,, \quad x = k\kappa a^{-2}\,.$$

The special case of solutions with $k=0$ admits a simple
four-dimensional interpretation. In this case the constant curvature
two-surface $\Sigma_2$ is simply $T^2$, so that the five-dimensional
metric can be trivially reduced to four dimensions by
 \be
\rd s_{\5}^2 = \rd s\4^2 + \rd y^2\,,
 \ee
with
 \be
\rd s\4^2 = g_{\alpha\beta}(x^{\gamma})\rd x^{\alpha}\rd x^{\beta} + \rd z^2\,.
 \ee
So solutions of Eqs. (\ref{nmg}) and (\ref{cons1}) with $k=0$ (if
they exist) will lead to cylindrical solutions of four-dimensional
quadratic gravity. Because the Gauss-Bonnet term is a topological
invariant in four dimensions, we do not expect the existence of such
solutions to depend on the value of the Gauss-Bonnet coupling
constant $\gamma$.

\setcounter{equation}{0}
\section{Solutions}
A number of exact solutions to the equations (\ref{nmg}) of NMG are
known \cite{bht,newmass,gaston,newmass2,bht2,gaston2,dynam}. Several
of these have a constant Ricci scalar. In the following we shall
concentrate on solutions of NMG with constant Ricci scalar, leading
to non-asymptotically flat black ring solutions of five- (or
possibly four-) dimensional quadratic gravity.

\subsection{BTZ}
Equations (\ref{nmg}) are trivially solved by Einstein metrics,
leading in the case of a negative Ricci scalar to BTZ black holes
\cite{bht,newmass}. The resulting five-dimensional solutions are
 \be\lb{bbtz}
\rd s_{\5}^2=-N^2 \rd t^2+\frac{\rd r^2}{N^2}+r^2\left(\rd z+N^z \rd
t\right)^2+a^2(\rd\theta^2 + s_k^2\rd\varphi^2)\,,
 \ee
where
 \be
N^2=\frac{r^2}{l^2}-M +\frac{J^2}{4r^2}\,,\quad N^z=\frac{J}{2r^2}.
 \ee
These solutions depend on two dynamical parameters $M$ and $J$
(integration constants), and two scale parameters $a^2$ ($\Sigma_2$
scale) and $l^2$ ($AdS_3$ scale). The $AdS_3$ scale is related to
the NMG parameters and to the Ricci scalar by \cite{bht,newmass}
 \be\lb{parabtz}
R = - 6l^{-2}\,, \quad l^{-2} = 2m^2\left[-1 \pm
\sqrt{1-\lambda/m^2}\right]\,.
 \ee
Inserting these in (\ref{paras}) and (\ref{cons2}), we obtain a
system of two quadratic equations for the rescaled parameters
$$x = k\kappa a^{-2}\,, \quad y = \kappa l^{-2}$$
in terms of the input parameters  $\alpha$, $\beta$, $\gamma$ and
$\ol\Lambda$:
 \ba
2\alpha x^2 - 4(\alpha+\gamma)xy - 2(3\alpha+13\beta)y^2 + x - y -
\ol\Lambda &=& 0\,, \nn\\
(4\alpha-\beta)x^2 - 12(\alpha+\gamma)xy - 72\beta y^2 + 3x - 6y -
4\ol\Lambda &=& 0\,.
 \ea
Eliminating $x$ between these equations generically yields an
equation of the sixth degree for the variable $y$. We consider here
only some special cases:

a)\underline{$\alpha=\beta=0$} (Gauss-Bonnet). In this case the
equations are solved by
 \be
x = \frac{2\La}{3+4\gamma\La}\,, \quad y = - \frac{\La}{3}\,.
 \ee
Assuming $\kappa > 0$, and comparing the signs of $x$ ($k$) and $y$
($l^{-2}$), we find that for $\gamma>0$, the five-dimensional
geometry is BTZ$\times S^2$ for $\La < -3/4\gamma$, BTZ$\times H^2$
for $-3/4\gamma<\La<0$, and $dS_3\times S^2$ for $\La > 0$, while
for $\gamma<0$ it is BTZ$\times H^2$ for $\La<0$, $dS_3\times S^2$
for $0<\La< -3/4\gamma$, and $dS_3\times H^2$ for $\La >
-3/4\gamma$. For $\gamma=0$ (pure Einstein case), the solution
reduces to the five-dimensional Nariai or anti-Nariai solution, with
the geometry $dS_3\times S^2$ for $\La
> 0$, or BTZ$\times H^2$ for $\La < 0$.

b) \underline{$\La =-(2\alpha+\beta)/(4\alpha+\beta)^2$}. In this
case,
 \be
x = - \frac1{4\alpha+\beta}\,, \quad y = 0\,,
 \ee
and the geometry is Minkowski$_3\times S^2$ or Minkowski$_3\times
H^2$ , depending on the sign of $4\alpha+\beta$. This can be
toroidally reduced to a solution of four-dimensional quadratic
gravity with the geometry Minkowski$_2\times \Sigma_2$.

c) \underline{$\La =-3(3\alpha+7\beta)/8(3\alpha+4\beta)^2$}.  This
leads to
 \be\lb{x0}
x = 0\,, \quad y = \frac1{3\alpha+4\beta}\,,
 \ee
with the geometry BTZ$\times T^2$ or $dS_3\times T^2$ according to
the sign of $3\alpha+4\beta$. This can also be toroidally reduced to
a solution of four-dimensional quadratic gravity, with the geometry
BTZ$\times S^1$ or $dS_3\times S^1$.

\subsection{AdS wave}
As shown in \cite{gaston} and \cite{newmass2}, the field equations
of NMG also admit AdS wave solutions which generalize extreme BTZ
black holes, and have the same constant Ricci scalar. From the
stationary solutions given in \cite{newmass2}, we obtain the
solutions of five-dimensional quadratic gravity
 \ba\lb{sd1}
\rd s_{\5}^2 &=& [-2l^{-2}\rho+ F(\rho)]\,\rd t^2 - 2lF(\rho)\,\rd
t\,\rd z + [2\rho + l^2F(\rho)]\,\rd z^2 \nn\\ && \quad +
\frac{l^2\,\rd \rho^2}{4\rho^2} +a^2(\rd\theta^2 +
s_k^2\rd\varphi^2)\,,
 \ea
with
 \be\lb{Fr}
F(\rho) = a_+\rho^{p_+} + a_-\rho^{p_-} + M/2\,, \quad p_{\pm} =
\frac{1\pm\sqrt{m^2l^2+1/2}}2\,,
 \ee
depending on three integration constants $a_+$, $a_-$ and the BTZ
mass parameter $M$. The $AdS_3$ scale $l^2$ is again given by the
BTZ relation (\ref{parabtz}) (the sign $\pm$ in (\ref{Fr}) is
independent from that in (\ref{parabtz})). For $m^2l^2 < -1/2$, the
two constants $a_{\pm}$ are complex conjugate, while for $m^2l^2 =
-1/2$ and $m^2l^2 = 1/2$, the form (\ref{Fr}) of $F(\rho)$
degenerates and must be replaced by forms involving logarithms,
which are given in \cite{newmass2}. For $a_+ = a_- = 0$, the
solution (\ref{sd1}) reduces, after the coordinate transformation
$\rho = r^2/2$, to the extreme BTZ solution (\ref{bbtz}) with
$J=Ml$. For $a_+ = 0$ but $a_- \neq 0$, the solution (\ref{sd1}) is
asymptotic (for $\rho \to \infty$) to the extreme BTZ solution for
$m^2l^2>1/2$, and weakly asymptotic (in the sense of log gravity) to
extreme BTZ for $m^2l^2=1/2$.

The analysis of \cite{newmass2} shows that the three-dimensional AdS
wave solutions describe regular black holes with a null Killing
vector, leading to regular five-dimensional black rings, only for
the discrete values $m^2l^2=17/2$, $m^2l^2=7/2$, and $m^2l^2=1/2$.
The black rings for $m^2l^2=17/2$ are actually a special case of the
warped AdS black rings discussed in the next subsection, so we only
consider the two other possibilities, focussing on the subcase
$k=0$. In this case, Eq. (\ref{x0}) gives $l^2 =
\kappa(3\alpha+4\beta)$, leading on account of (\ref{paras}) to
 \be
m^2l^2 = \frac{18\alpha-11\beta}{2\beta}\,.
 \ee
The massless $m^2l^2=7/2$ black rings are obtained for
$\alpha=\beta$, $\La=-15/196\beta$. The five-dimensional metric is
given by (\ref{sd1}), (\ref{Fr}) with $k=0$, $a_-=M=0$, $a_+ > 0$,
$p_+ = 3/2$, with a double horizon at $x\equiv \rho^{1/2}=0$ hiding
a timelike causal singularity (for details, see \cite{newmass2}).

If the coupling constants are related by $\alpha=2\beta/3$,
$\La=-3/32\beta$, the choice $k=0$ leads to $m^2l^2=1/2$. The
logarithmic solution which replaces (\ref{Fr}) in this case leads to
three possible kinds of regular five-dimensional black rings (or
four-dimensional black strings). 0nly one of which is massive. with
the four-dimensional black string metric (after an appropriate
coordinate transformation)
 \ba\lb{r213}
\rd s_{(4)}^2 &=& -\frac{4\rho^2}{l^2r^2}\rd t^2 + r^2\bigg[
\rd\varphi - \frac{bl\ln|\rho/\rho_0|}{r^2}\rd t\bigg]^2 +
\frac{l^2\rd\rho^2}{4\rho^2} + \rd z^2 \nn\\ && \qquad (r^2 = 2\rho
+ bl^2\ln|\rho/\rho_0|)\,,
 \ea
depending on two parameters $b<0$ and $\rho_0>0$. Again, the double
horizon at $\rho=0$ shields a timelike causal singularity. This
spacetime is asymptotically $AdS_3\times S_1$ in the sense of log
gravity, and its mass and angular momentum satisfy the extremality
condition $J = Ml$. That this mass, given by
 \be\lb{masslog}
M = \frac{2b\tau}{G_3}
 \ee
(with $G_3$ the effective three-dimensional Newton constant and
$\tau$ the period of the coordinate $z$), is positive is non
trivial. As discussed in \cite{newmass2}, $b$ must be negative in
order to avoid naked CTC, so that for a positive $G_3$, $M$ is
negative. However, comparing (\ref{eom3}) and (\ref{paras}), we see
that the effective three-dimensional Newton constant
 \be
G_3 \propto c^{-1} = - \frac1{\beta m^2} = - \frac{\kappa}{\beta
y\,m^2l^2} = -12\kappa
 \ee
(using (\ref{x0}) with $\alpha=2\beta/3$ and $m^2l^2=1/2$) is
negative definite, ensuring a positive mass (\ref{masslog}).

The stationary AdS wave solutions of \cite{newmass2} are actually
special cases of the more general $AdS_3$ wave solutions of NMG
\cite{gaston}, which similarly lead to solutions of five-dimensional
quadratic gravity
 \be\lb{sd2}
\rd s_{\5}^2 = 2l^{-2}\rho\,\rd u\,\rd v + F(\rho,u)\,\rd u^2 +
\frac{l^2\,\rd \rho^2}{4\rho^2} +a^2(\rd\theta^2 +
s_k^2\rd\varphi^2)\,,
 \ee
where $u=lz-t$, $v=lz+t$, and $F(\rho,u)$ is given by (\ref{Fr})
with the constants $a_{\pm}$ replaced by arbitrary functions of $u$.
These are different from the $AdS_5$ wave solutions of \cite{ggst}.
Another special case of AdS wave solutions of NMG is the dynamical
black hole metric of \cite{dynam}, which can similarly be uplifted
to five dimensions.

\subsection{Warped AdS}
The warped $AdS_3$ solutions of NMG (previously discussed as
solutions of topologically massive gravity \cite{djt} (TMG)
\cite{adtmg,ALPSS,tmgebh}) given in \cite{newmass} lead to the
following solutions of five-dimensional quadratic gravity:
 \ba \rd s_{\5}^2 &=& -b^2\frac{\rho^2-\rho_0^2}{r^2}\,\rd t^2 +
r^2\bigg[\rd z - \frac{\rho+(1-b^2)\omega}{r^2}\,\rd t\bigg]^2 \nonumber \\
&&\qquad + \frac1{b^2\z^2}\frac{\rd\rho^2}{\rho^2-\rho_0^2}+
a^2(\rd\theta^2 + s_k^2\rd\varphi^2)\,, \lb{warp}
 \ea
where
 \be\lb{r2} r^2 = \rho^2 +2\omega\rho + \omega^2\,(1-b^2) +
\frac{b^2\rho_0^2}{1-b^2} \,,
 \ee
and the constants $b^2$ and $\z$ are given by
 \be\lb{beta2} b^2 = \frac{9 - 21\lambda/m^2 \mp
2\sqrt{3(5+7\lambda/m^2)}}{4(1-\lambda/m^2)}\,, \quad \z^{-2} =
\frac{21-4b^2}{8m^2}\,.
 \ee
As shown in \cite{tmgebh}, these metrics correspond to regular black
holes provided
 \be\lb{reg}
\zeta^2>0\,, \quad 0<b^2<1\,, \quad \omega>0\,.
 \ee
In the limiting cases $b^2 = 1$ and $b^2 = 0$, the three-dimensional
reduced metric should be replaced by the special solutions given in
\cite{tmgebh} (where $\mu_E$ should be replaced by $\z$).

We recall that the NMG parameters $\lambda$ and $m^2$ in
(\ref{beta2}) are related to the original five-dimensional
parameters and to the three-dimensional curvature associated with
(\ref{warp}),
 \be\lb{Rwarp}
R = \frac{\z^2}2\,(1-4b^2)
 \ee
by (\ref{paras}) and the constraint (\ref{cons2}). The general case
is intricate and unlightening, so we concentrate on the special case
$\alpha=\gamma=0$. To compute the values of the solution parameters
corresponding to a given set of values of the model parameters, we
proceed in the following fashion. Eq. (\ref{beta2}) can be inverted
to yield, for each value of $b^2$, two values of the ratio
$\lambda/m^2$:
 \be\lb{lm2ab}
a)\;\; \frac{\lambda}{m^2} = \frac{16b^4-72b^2+21}{(4b^2-21)^2}\,,
\quad b)\;\; \frac{\lambda}{m^2} = 1
 \ee
(for $\lambda/m^2=1$ the first equation (\ref{beta2}) with the lower
sign is indeterminate\footnote{This fact was overlooked in
\cite{newmass}.}). From (\ref{Rwarp}) and the second and third Eqs.
(\ref{paras}), we obtain
 \be\lb{warpac0}
\beta\ol{R} = \frac{4b^2-1}5\,, \quad \z^{-2} =
-\frac52\beta\kappa\,.
 \ee
Using this together with the Eqs. (\ref{paras}), we can compute
another value of the ratio $\lambda/m^2$ in terms of the solution
parameters $b^2$ and $x$:
 \be
\frac{\lambda}{m^2} =
\frac{400\beta(x-\La)-11(4b^2-1)^2}{(4b^2-21)^2}\,.
 \ee
Comparing this with (\ref{lm2ab}), we obtain for each value of $b^2$
two possible values for $\beta(x-\La)$:
 \be\lb{xLa}
a)\; \beta(x-\La) = \frac{2(2b^2-1)(3b^2-1)}{25}\,, \quad b)\;
\beta(x-\La) = \frac{48b^4-64b^2+113}{100}\,.
 \ee
Finally these results are combined with the constraint
(\ref{cons2}), written as
 \be
(\beta x + 1/2)^2 = 4\beta(x-\La) + 3/8 - 2(\beta\ol{R}-1/4)^2\,,
 \ee
to obtain the two possible relations between the solution parameters
$x = k\kappa a^{-2}$ and $b^2$:
 \be\lb{xb2}
a) \beta x = - \frac12 \pm \frac{\sqrt{64b^4-16b^2+29}}{10}\,, \quad
b) \beta x = - \frac12 \pm \frac{\sqrt{64b^4-112b^2+449}}{10}\,.
 \ee
The corresponding values of the model parameter $\beta\La$ are given
in terms of $b^2$ by (\ref{xLa}) and (\ref{xb2}). The sign $k$ of
the two-dimensional curvature is determined from the sign of $\beta
x$, taking into account that $\beta\kappa < 0$ from the second
equation (\ref{warpac0}). The outcome is that for both cases a) and
b) in (\ref{xb2}), $k= \mp1$, so that there is no solution of this
kind in four-dimensional quadratic gravity ($k=0$). The mass of the
five-dimensional warped AdS black rings is proportional to the
three-dimensional mass of warped AdS black holes in NMG, which was
computed in \cite{newmass} to be
 \be\lb{Mwbh}
M = \frac{\zeta^3b^2(1-b^2)}{2G_3m^2}\omega \propto
b^2(1-b^2)\omega\,,
 \ee
 where we have used (\ref{paras}) and
(\ref{warpac0}). This mass is positive by virtue of the regularity
constraints (\ref{reg}).

\subsection{$AdS_2\times S^1\times\Sigma_2$}
In the special case $\lambda=m^2$, another solution of NMG is
$AdS_2\times S^1$ \cite{newmass}, leading to the five-dimensional
solution
 \be\lb{BR1}
\rd s^2 = -(\rho^2-\rho_0^2)\rd t^2 - 2m^2\rd z^2 -
\frac{\rd\rho^2}{2m^2 (\rho^2-\rho_0^2)} + a^2(\rd\theta^2 +
s_k^2\rd\varphi^2)\,.
 \ee
This solution -- which leads to a four-dimensional solution after
integrating out the cyclic coordinate $z$ -- has the Minkowkian
signature in the range $\rho^2>\rho_0^2$ provided $m^2<0$ and
$a^2>0$. The three-dimensional curvature is $R=4m^2$. Inserting this
into the second and third equations (\ref{paras}), we obtain the
linear relation
 \be\lb{lin}
1 + 4(\alpha+\gamma)x + 4\alpha y = 0
 \ee
between the rescaled parameters
 \be
x = \kappa ka^{-2}\,, \quad y = 2\kappa m^{2}\,.
 \ee
This relation shows that there is no solution of this kind for
$\alpha=\gamma=0$.

One can show that there is also no solution for $\alpha=0$, even if
$\gamma\neq0$. For $\alpha\neq0$, Eq. (\ref{lin}) can be used to
eliminate $y$ from the first equation (\ref{paras}) with
$\lambda=m^2$ and the constraint (\ref{cons2}), leading to the
system
 \ba\lb{sys}
&&\left[\alpha^2(3\beta+4\gamma) + 2\alpha\gamma(3\beta+\gamma) +
3\beta\gamma^2\right]x^2 + \left[\frac{\alpha}2(3\beta + 2\gamma) +
\frac{3\beta\gamma}2\right]x \nn\\ && \qquad +
\frac{\alpha}8 + \frac{3\beta}{16} + \alpha^2\ol\Lambda = 0\,, \nn\\
&& \left[\alpha^2(9\beta+8\gamma) + 4\alpha\gamma(4\beta+\gamma) +
8\beta\gamma^2\right]x^2 + \left[\alpha(4\beta + 3\gamma) +
4\beta\gamma\right]x \nn\\ && \qquad + \frac{\alpha+\beta}{2} +
4\alpha^2\ol\Lambda = 0\,,
 \ea
which is overdetermined, meaning that there must be a specific
relation between the model parameters $\alpha$, $\beta$, $\gamma$,
and $\ol\Lambda$.

Again the general case is intricate, so we focus on the example
$\gamma=0$. In this case the system (\ref{sys}) has a solution
provided the model parameters are related by
 \be
3\xi^2 + 2(6\alpha-11\beta)\xi + 12\alpha^2+20\alpha\beta + 3\beta^2
= 0,\,,
 \ee
where $\xi=16\alpha^2\La$. The further assumption $\Lambda=0$ leads
to the two possibilities
 \ba
{\mathrm a}) & 2\alpha + 3\beta = 0 \,, & \quad y = -\frac{x}2 =
-\frac1{6\beta}\,, \\
{\mathrm b}) & 6\alpha + \beta = 0 \,, & \quad y = \frac{x}2 =
\frac1{2\beta}\,.
 \ea
Remembering that $m^2 < 0$, we find that in the first case, $k=+1$
and the five-dimensional geometry is $AdS_2\times S^2\times S^1$,
while in the second case, $k=-1$ and the geometry is $AdS_2\times
H^2\times S^1$.

Returning to the general case, let us point out that, contrary to
appearance, these five-dimensional solutions with the geometry
$AdS_2\times S^1\times\Sigma_2$ will not generically lead, upon
integration along the $S^1$, to Bertotti-Robinson-like solutions of
four-dimensional quadratic gravity. The reason is that the
five-dimensional equations (\ref{eom}) contain, besides the
equations of four-dimensional quadratic gravity, the $(zz)$
component which leads to the additional constraint ${\cal L}_4 = 0$,
where ${\cal L}_4$ is the integrand of the quadratic action
(\ref{action}) for $D=4$. However, as already mentioned, the case
$k=0$ does lead upon toroidal reduction to a solution o
four-dimensional quadratic gravity. Assuming $x=0$, we obtain from
(\ref{lin}) $y = -1/4\alpha$, which satisfies the system (\ref{sys})
provided
 \be\lb{at}
\beta=0\,, \quad \La = -1/8\alpha\,.
 \ee
Note that $\beta=0$ with $m^2$ finite means from (\ref{paras}) that
$c=0$, contrary to the assumption made in deriving (\ref{nmg}) from
(\ref{eom3}). However it can be checked directly that equations
(\ref{eom3}) are satisfied for these relations between the coupling
constants if $R = - 1/2\kappa\alpha$. It is tempting to speculate
that this $AdS_2\times S^1\times S^1$ solution could be the
near-horizon limit of some four-dimensional black ring solution to
$f(R)$ gravity.

\section{Conclusion}

In this paper, we have constructed new solutions of five-dimensional
quadratic gravity as direct products $M_3\times\Sigma_2$, where
$\Sigma_2= S^2$, $T^2$ or $H^2$ is a constant curvature two-surface,
and $M_3$ is a solution of three-dimensional new massive gravity
with constant scalar curvature. These non-asymptotically flat
solutions could represent near-horizon limits of five-dimensional
asymptotically flat black strings or black rings. A number of these
solutions are themselves black strings or rings (topological if
$\Sigma_2= T^2$ or $H^2$) of the BTZ (\ref{bbtz}), null Killing
vector (\ref{sd1}) or warped $AdS_3$ (\ref{warp}) type.

A by-product of our analysis is the construction of new solutions of
four-dimensional quadratic gravity with the geometry $M_3\times
S^1$, obtained by toroidal reduction of the five-dimensional
solutions with flat transverse space. These again include black
strings or rings of the BTZ or null Killing vector type (among which
the log black string (\ref{sd2})), and an $AdS_2\times T^2$ solution
of $f(R)$ gravity for a specific relation between the model
parameters.

We close by commenting on the relation of the present work with
recent work on higher-dimensional supergravities with curvature
squared invariants \cite{lps,brs,bcsp,ozpa}. In \cite{lps}, a
six-dimensional supergravity with quadratic couplings was reduced on
a three-sphere, yielding a massive three-dimensional supergravity,
which is closely related to the general massive supergravity of
\cite{bht2}. Freezing out the scalars in this theory leads to the
general massive gravity (NMG + TMG) of \cite{bht}. Similarly to what
has been done here, we expect solutions of the latter theory
\cite{gmt1,gmt2} to lead to Bertotti-Robinson-like solutions of the
original six-dimensional theory. The five-dimensional supergravity
of \cite{brs} with a Riemann squared invariant is expected to
compactify over a two-sphere to a supersymmetric extension of
topologically massive gravity. Likewise, known solutions of TMG
should lift to solutions of the five-dimensional theory. In
\cite{bcsp} a higher derivative extension of a six-dimensional
gauged supergravity was shown to admit solutions given by direct
products $M_4\times\Sigma_2$ or $M_3\times\Sigma_3$, with $M_p$ and
$\Sigma_q$ both constant curvature. In \cite{ozpa} a supersymmetric
completion of the Einstein-Gauss-Bonnet theory was similarly shown
to admit solutions with $AdS_3\times S^2$ and $AdS_2\times S^3$
structures. Presumably these theories should also admit more general
product solutions similar to those presented here.

\subsection*{Acknowledgment}
I wish to thank Dmitry Gal'tsov for stimulating discussions.

\end{document}